\documentclass[twocolumn]{aastex631}
\setcitestyle{super, comma, numbers}

\usepackage{xcolor}       
\usepackage{soul}
\usepackage{ulem}

\begin{document}



\shorttitle{Superlarge pebbles}
\shortauthors{Teiser et al.}

\title{The growth of super-large pre-planetary pebbles to an impact erosion limit}

\author[0000-0003-4468-4937]{J. Teiser}
\affiliation{University of Duisburg-Essen, Faculty of Physics,
Lotharstr. 1, 47057 Duisburg, Germany}

\author[0009-0003-9097-1934]{J. Penner}
\affiliation{University of Duisburg-Essen, Faculty of Physics, Lotharstr. 1, 47057 Duisburg, Germany}

\author[0000-0002-5841-2636]{K. Joeris}
\affiliation{University of Duisburg-Essen, Faculty of Physics, Lotharstr. 1, 47057 Duisburg, Germany}

\author[0000-0001-8486-4743]{F. C. Onyeagusi}
\affiliation{University of Duisburg-Essen, Faculty of Physics,
Lotharstr. 1, 47057 Duisburg, Germany}

\author[0000-0002-3517-1139]{J. E. Kollmer}
\affiliation{University of Duisburg-Essen, Faculty of Physics,
Lotharstr. 1, 47057 Duisburg, Germany}

\author{D. Daab}
\affiliation{Swedish Space Corporation (during experiment design and flight)}

\author[0000-0002-7962-4961]{G. Wurm}
\affiliation{University of Duisburg-Essen, Faculty of Physics,
Lotharstr. 1, 47057 Duisburg, Germany}


\begin{abstract}
Early dust evolution in protoplanetary disks is dominated by sticking collisions. However, this initial phase of particle growth faces constraints - notably from destructive encounters. To find the maximum particle size achievable, we studied collisional processes during a prolonged microgravity experiment aboard a suborbital flight. Here, we specifically report an impact erosion limit. We observed individual basalt beads, each measuring 0.5 mm in diameter, colliding with and either eroding or adhering to a cluster several centimeters in size. This cluster, formed from tribocharged particles, simulates an electrostatic growth phase that surpasses the classical bouncing barrier. We find a threshold velocity of about 0.5 m/s, distinguishing between additive and erosive impacts of individual beads. Numerical simulations of grain impacts into clusters, testing both low and high charge constituents corroborate
the experimental findings of surface erosion within the observed velocity range.
This specific velocity threshold suggests the potential formation of pebbles several centimeters in size within protoplanetary disks. Such dimensions place these pebbles well into a regime where hydrodynamic interaction might facilitate the formation of planetesimals.
\end{abstract}

\section{Note on Publication}
This version of the article has been
accepted for publication, after peer review but is not the Version of Record and does not
reflect post-acceptance improvements, or any corrections. The Version of Record is available (open access) online at:
https://www.nature.com/articles/s41550-024-02470-x

\section{Main}

In recent years, a class of objects known as \textit{pebbles} got increasing attention in the study of planet formation.
Unlike in geology, the precise size of a pebble remains elusive within the field. It is often associated with a Stokes number of $\rm St \sim 1$ indicating a particle that couples to the gas of a protoplanetary disk in a single orbit.
For compact particles, this size range extends from several centimeters to a decimeter within the inner few AU of a disk \citep{Johansen2014}.

Pebbles play a crucial role as they serve as a reservoir of particles throughout the lifespan of protoplanetary disks, available for accretion. Pebble accretion is proposed for the efficient growth of terrestrial planets and Super-Earths \citep{Savvidou2023, Demirci2020accretion, Vazan2023}. In the earlier stages of evolution, streaming instabilities may gather pebbles into dense clumps that could eventually collapse into planetesimals \citep{Youdin2005, Johansen2006,  Gerbig2020}. 
It is conceivable that such pebbles persist over extended periods, given that collisional growth faces bouncing or fragmentation barriers 
\citep{Guettler2010, Zsom2010, Wurm2021nat, Birnstiel2011, Drazkowska2021}.
Just the precise limiting size remains uncertain.

From a numerical or theoretical standpoint, certain Stokes numbers are deemed necessary owing to the hydrodynamic behavior, independent of the mechanical properties and collisional outcomes. Generally,  centimeter-sized particles might be required \citep{Carrera2022, Drazkowska2014}. 
However, from an experimental perspective, millimeters are typically the largest size achievable through hit-and-stick in a standard manner, before aggregates encounter a bouncing limit \citep{Zsom2010, Kelling2014, Kruss2016, Demirci2019b}. 
{Also, recent numerical simulations of collisions converge on this picture that aggregates at the bouncing barrier are rather small \citep{Arakawa2023}.}

Various conditions exist to locally increase aggregate size under stickier conditions, such as elevated temperatures resulting in water loss or the presence of magnetic fields for particles rich in metallic iron \citep{Pillich2021, Kruss2018, Bogdan2023}.
One of the most promising approaches though involves clustering of tribocharged aggregates. This suggests a novel growth phase beyond the bouncing barrier. As particles collide, bounce off each other and charge alongside they can assemble into larger clusters of aggregates, eventually strongly bound by electrostatic forces. Recent studies indicate the potential of this mechanism for the formation of large clusters \citep{Steinpilz2020, Jungmann2018, Jungmann&Wurm2021, Teiser2021, Lee2015}. Despite their differing morphology compared to large compact dust aggregates, hydrodynamic interactions would perceive these clusters as large particles, regardless of composition.

A lingering question pertains to the maximum size attainable by clusters under disk conditions until they encounter the next growth barrier, whether it be cluster bouncing, cluster fragmentation or cluster erosion. These limits remain elusive.
Here, we focus on the erosion limit.

Erosion is defined as the gradual removal of surface material from a larger body, in our case, via impacts from individual (sub)-millimeter grains onto a larger cluster of these grains.  

We aim to determine the maximum size of a particle cluster capable of withstanding such bombardment in protoplanetary disks. Keeping in mind that there will be some transitional range, the core query revolves around the impact speed at which individual grains induce cluster erosion. Upon reaching this limit, more particles are liberated, becoming further projectiles for aggregates, perpetuating erosion until the aggregates diminish in size sufficiently for collision velocities to fall below the erosion threshold.

\subsection{Suborbital experiment}

To study this in experiments, grains need to be charged, clusters of these grains need to be built and these clusters have to be subjected to impact with individual grains at varying speeds. Microgravity is imperative for studying free collisions and charged clusters cannot form undisturbed on the ground. The entire process requires minutes to execute, thus we conducted a suborbital experiment. A sketch of the experiment is shown in the supplementary figure 1. 
The sample studied comprises essentially identical, monodisperse basalt beads with a diameter of 0.5 mm.
Details on the experimental procedures can be found in Methods.

Upon entering microgravity, the particle reservoir is opened, and the particles are gently released through moderate shaking. Through this process, particles collide gently and gradually assemble into one big aggregate, as depicted in fig. 1. 
This marks the starting point of our investigations herein.
\begin{figure*}
	\centering
\includegraphics[width=\linewidth]{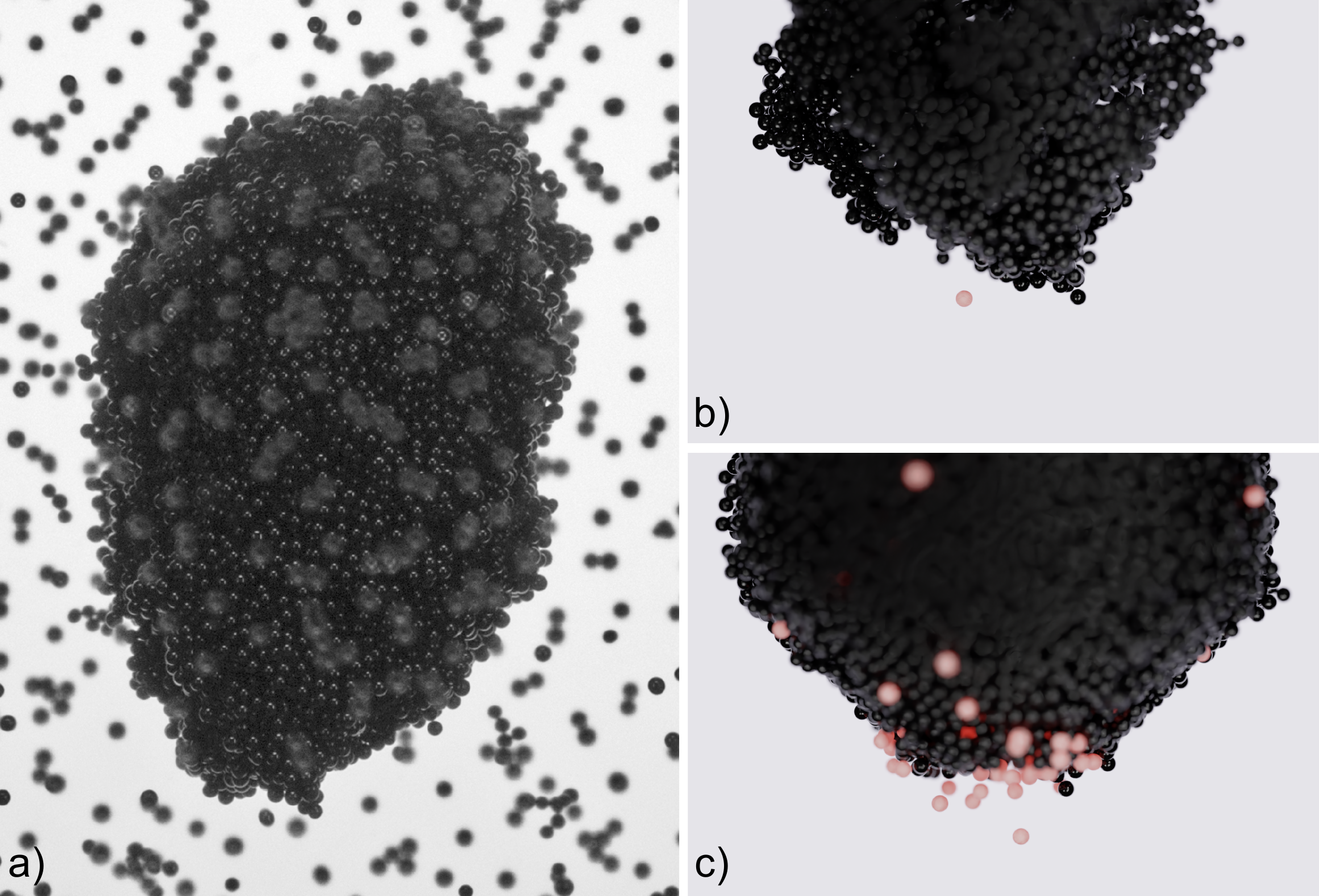}
\caption{\label{giang} a) Largest cluster formed in the experiment of about 3\,cm in length. Unfocused grains are particles at the front window. For reference, individual grains are 0.5\,mm. b) and c) Numerical simulation of an impact at 0.5\,m/s. Impactor and ejected grains are marked in red. c) With weak charge some grains are ejected. b) With strong charge no particles are ejected. Only the impactor is seen.} 
\end{figure*}

\begin{figure}
	\centering

\includegraphics[width=\columnwidth]{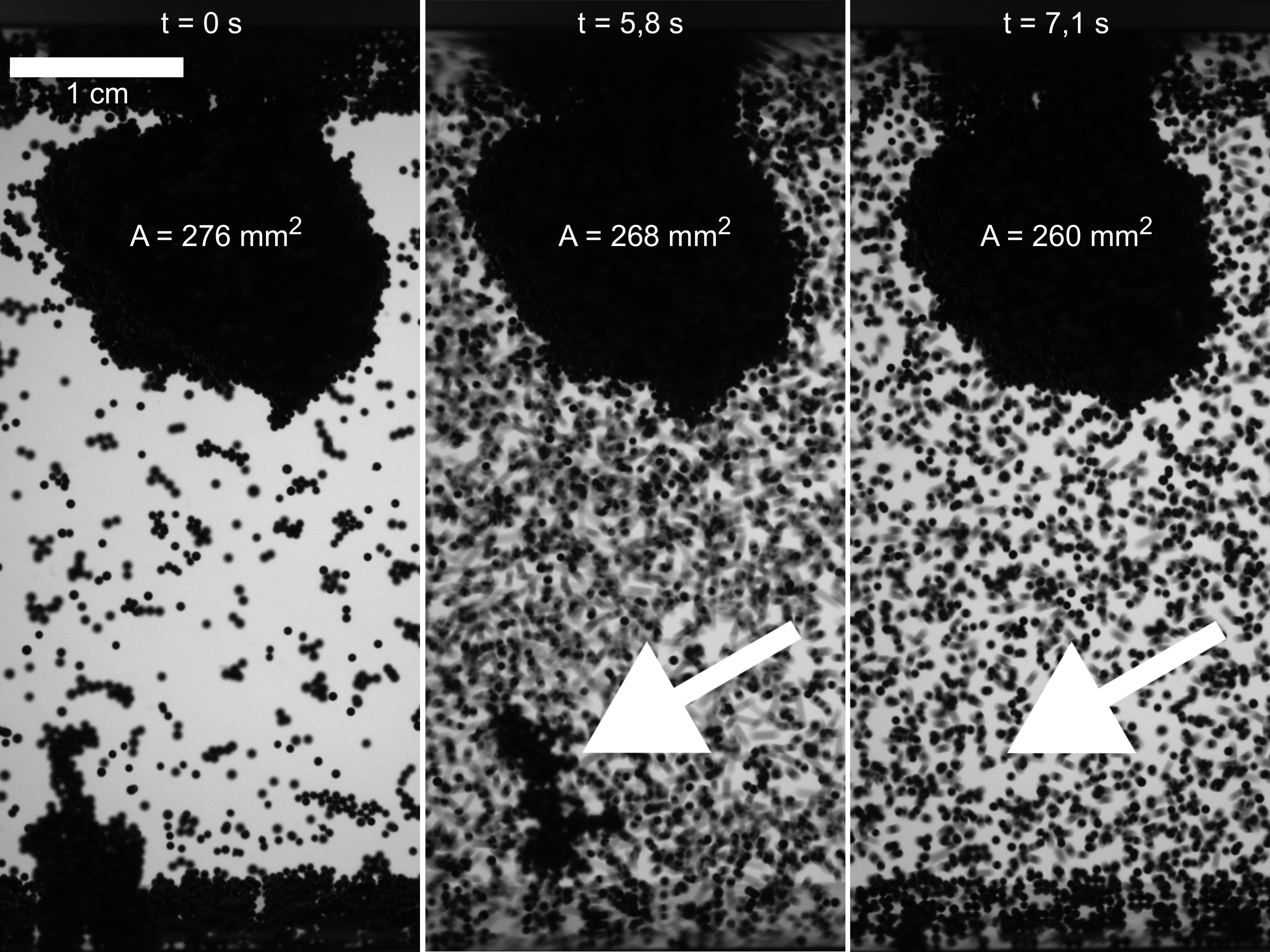}
\caption{\label{erosion} Time series of images of eroded basalt clusters. Shown is the shrinking of the clusters as they are eroded by impacts of individual grains. We marked a smaller cluster which vanishes completely in the course of a few seconds. {Imprinted in the large cluster is its cross section decreasing with time.}} 
\end{figure}

\section{Results}

The experiment chamber underwent shaking of various degree.
As the granular gas heats up, collisions with the large cluster become fast enough to initiate surface erosion, causing grains to detach from its outer surface.
Fig. 2

presents a series of images taken at subsequent times of the eroding cluster. 
The shrinking of the cluster can be measured from this {and its decreasing area is imprinted within the figures}, but erosion is particularly evident on a smaller cluster initially present. Due to its smaller size, the volume reduction is more efficient, leading to its complete disappearance within a few seconds.

\begin{figure*}
	\centering
\includegraphics[width=\textwidth]{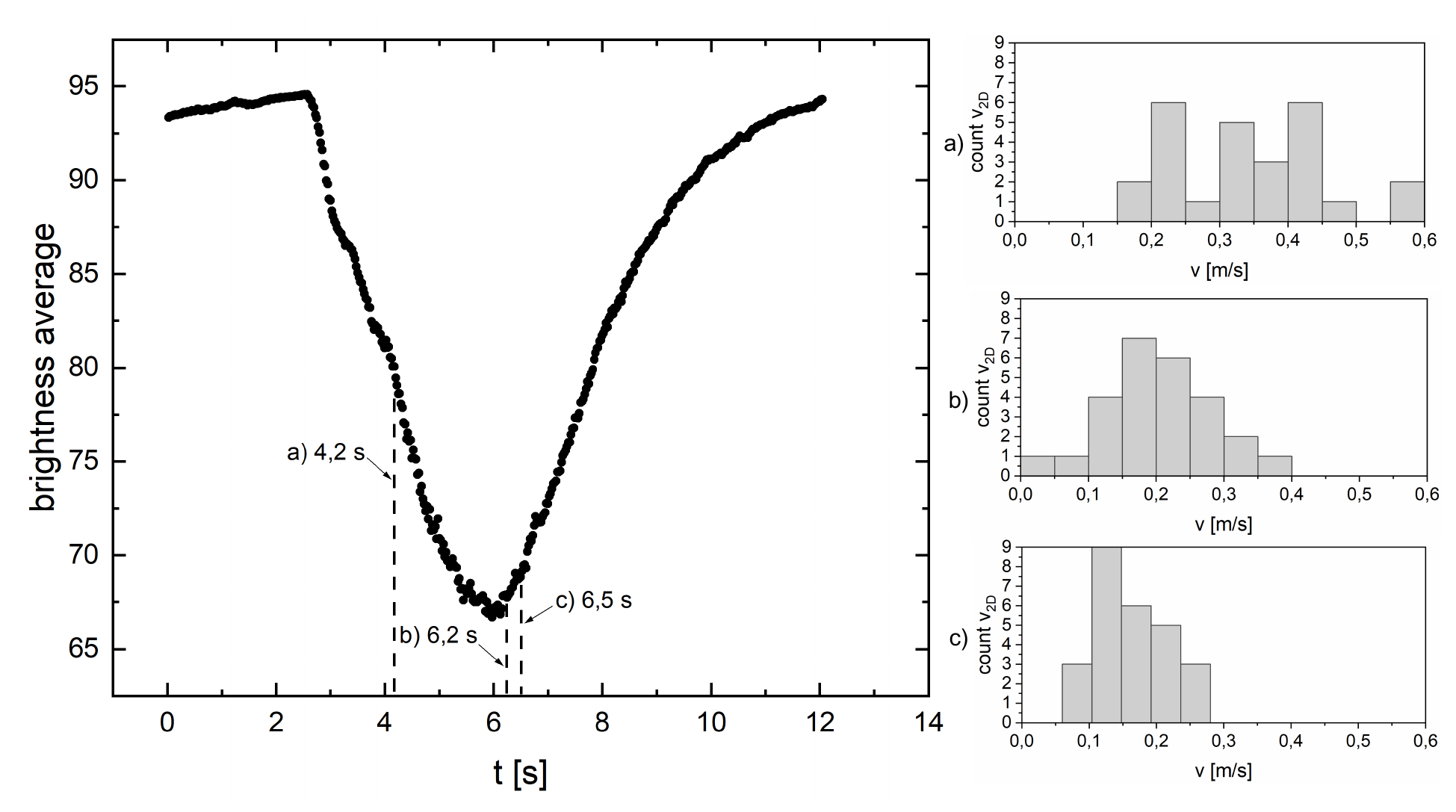}
\caption{\label{bright} Average pixel brightness of the images over time. {Total range would be 0 (black) to 255 (white).} Vibrations start at about 2 to 3\,s and stop at about 6\,s. Marked are times when the velocity distributions on the right are taken. Times and distributions are marked by letters. 
(a) velocities at the onset of erosion (t~=~4.2\,s). (b) velocities at end of vibrations (t~=~6.2\,s). {Erosion is no longer the dominant process. Therefore, only the largest velocities in (a) and (b) are indicative of the erosion threshold. (c) velocities when the dominant collisions are no longer erosive but small aggregates re-form (t~=~6.5\,s).}} 
\end{figure*}

Erosion alters the average brightness of the images \citep{Schneider2021}.
In our case, the onset of erosion is clearly identifiable by changes in brightness. As aggregates disintegrate and individual grains become more abundant, the images progressively become more opaque, as depicted in fig. 3.

The erosion threshold is defined as the minimum speed at which erosion occurs. We determine this threshold by measuring the velocity of individual particles during erosion.
Specifically, we derive speed from the track lengths of free particles over exposure time. 
We analyze velocity distributions at two distinct stages: when erosion is well underway and when it is about to cease.  
These distributions are also depicted in fig. 3.

Times at which velocities are measured are marked by lines and letters (a) and (b) in the brightness plot in fig. 3.  
At t~=~6.2\,s the brightness starts to increase again which implies that erosion is no longer dominant. This implies that the highest velocities are indicative of the threshold erosion velocity.

To establish the lower limit, we collect additional velocity data at the point when erosion transitions back into aggregation. 
Once the chamber motion ceases and relaxation slows down particles, brightness increases again, and small clusters begin to reform (not shown).
{Also the remains of the larger cluster then start to collect grains again and regrow.}
The velocity distribution at this time is also represented as a histogram in fig. 3
(c). 
In comparison to erosion, we find that speeds higher than 0.3\,m/s are absent. 

{That suggests collisional outcomes roughly align to no erosion occuring below 0.4\,m/s and erosion occuring at minimum at 0.6\,m/s} {or} $v_e = 0.5 \rm \pm 0.1 \, m/s$ as the erosion threshold.
All analyzed velocities are larger than 0.1 m/s at the onset of re-aggregation. This implies that the threshold between sticking and bouncing is larger than this and 0.1 m/s is a lower limit.
It is essential to note that thresholds may vary under different experimental conditions.

\subsection{Numerical simulations}

The observed cluster is surrounded by a cloud of particles and each impact liberates a number of grains locally near the impact point. To supplement our observations, we conducted numerical impact simulations under conditions closely resembling the experiments. Details are given in  Methods.
To replicate experimental conditions, we employed an impact velocity of 0.5\,m/s and observed erosion with weakly charged particles, ejecting a handful of grains. Conversely, strongly charged clusters remained intact at the same velocities and even at velocities exceeding 2.5\,m/s. A snapshot of an impact is shown in fig. 
1, with weak (c) and strong (b) charge.

While numerical simulations do not precisely replicate experimental conditions, they provide valuable insights into the role of charge in erosion dynamics. It also has to be noted that the charge is usually not distributed homogeneously on the surface as e.g. seen in measured dipole moments \citep{Steinpilz2020a, Onyeagusi2022}. So even overall neutral grains can provide strong binding if two spots of opposite polarity on two grains are close to each other. {This will also support sticking in cluster-cluster collisions.} The inclusion of this is beyond the scope of this paper.

However, with similar parameters as in the experiments, and few or no particles being liberated at a relevant velocity and charge, the numerical simulations support the idea that a high erosion threshold of a cluster is set by the charges on the constituents.

\section{Discussion}

The measured erosion threshold has to be set in the context of planet formation. It corresponds to the collisions with a cluster of several cm in size \cite{Weidenschilling1993}. That means that the cluster shown in fig.
1 is actually close to what has to be expected.
{Some meteorites largely consist of sub-mm sized chondrules which are currently suggested to gather in some particle traps \citep{Gurrutxaga2024}. In this case, our experiments with solid, monolithic grains are directly applicable. The clusters in our experiments and simulations have an average porosity between 50 to 60\%.}
Outside of chondrule regions, initially clusters in disks will consist of dust aggregates rather than monolithic grains and the size and material might be somewhat variable. 
{From recent drop tower experiments, we know that dust aggregates can charge to the same order of magnitude and this charge can also promote cluster growth at speeds beyond regular bouncing (Onyeagusi et al. submitted).}

The detailed influence of parameter variations is currently somewhat speculative but there might be trends. Smaller constituents tend to form more ruggedized charged clusters. This is what test measurements leading to the experiments reported here showed where 0.2\,mm basalt particles were used instead of 0.5\,mm. There, cm-aggregates were incredibly hard to shatter at all though we did not quantify this further. 
There might be more variation coming with material or temperature, so there is no final cluster size limit that can be given for every spot within the disk, yet.

{Erosion as limiting factor has also been studied numerically, e.g. by \cite{Krijt2015} or \cite{Seizinger2013} and compared to experiments in \cite{Schraepler2018}. These simulations were for very small projectiles on dust grain level and found erosion thresholds more on the order of 10 m/s. This is way beyond the phase we consider here but will be relevant for collisions in later phases of planetesimal and disk evolution.}

\subsection{Charging versus discharging}

The application to protoplanetary disks asks for the persistence of electric charges on dust aggregates. This cannot be taken for granted. This is less of a topic for laboratory experiments. The collisional timescales in the laboratory or suborbital flight are only minutes. This is quite different from collisional timescales in protoplanetary disks. In disks they might regularly collide on a timescale of years or less \citep{Dominik2024}. Trapping, e.g. in pressure maxima or vortices, with higher particle densities might reduce this timescale to months, weeks or less. In any case, this is a long time. If the dust aggregates are conductive enough, and small conductivity should do, then the different polarities might cancel out on these timescales, eventually. This would leave a neutral aggregate or cluster behind. The latter might then not have formed in the first place though. {Also, particles in a (weakly) ionized disk could be charged and discharged by electrons and ions. At the surface of protoplanetary disks, this can prevent growth \citep{Okuzumi2009, Akimkin2020}. We do not consider this to be important in the midplane. To quantify this, we} also carried out measurements on internal and external discharge \citep{Becker2024_discharge}. \citep{Becker2024_discharge} find that timescales between weeks and years are feasible for charges to stay on the particles. Dust aggregates should therefore be capable of holding on to their charges long enough. 

\subsection{Conclusion}

From the current suborbital experiment, we obtained an initial estimate for a limit in cluster growth of charged particles, likely ranging from centimeters to a decimeter. Therefore, the cluster depicted in fig. 1 
could serve as a close representation of the largest clusters in protoplanetary disks. Erosion occurs at approximately 0.5\,m/s with the given sample. Interestingly, this is not significantly different from the lower limit sometimes assumed as a size-independent value for fragmentation of large dust aggregates \citep{Birnstiel2011, Drazkowska2021}. While collisions between aggregates of equal size might yield different results, they might not necessarily lead to much lower fragmentation speeds. 

Notably, this marks the first instance where such high destructive threshold speeds are measured for particles of centimeter-size. It is crucial to highlight that growth solely from dust without charge would not permit the formation of such large entities, as clusters would be dismantled at speeds of millimeters per second \citep{Kelling2014}. Charged aggregation presents a feasible explanation for previously assumed large fragmentation speeds, although the presence of charge may introduce additional complexities in collisional growth dynamics.

Within an erosion limit, particles have the potential to bounce, acquire charge, and cluster into sizes susceptible to capture by streaming instabilities, eventually leading to gravitational collapse into planetesimals. This view is schematically visualized in fig. 4. 

\begin{figure*}
	\centering
    \includegraphics[width=\textwidth]{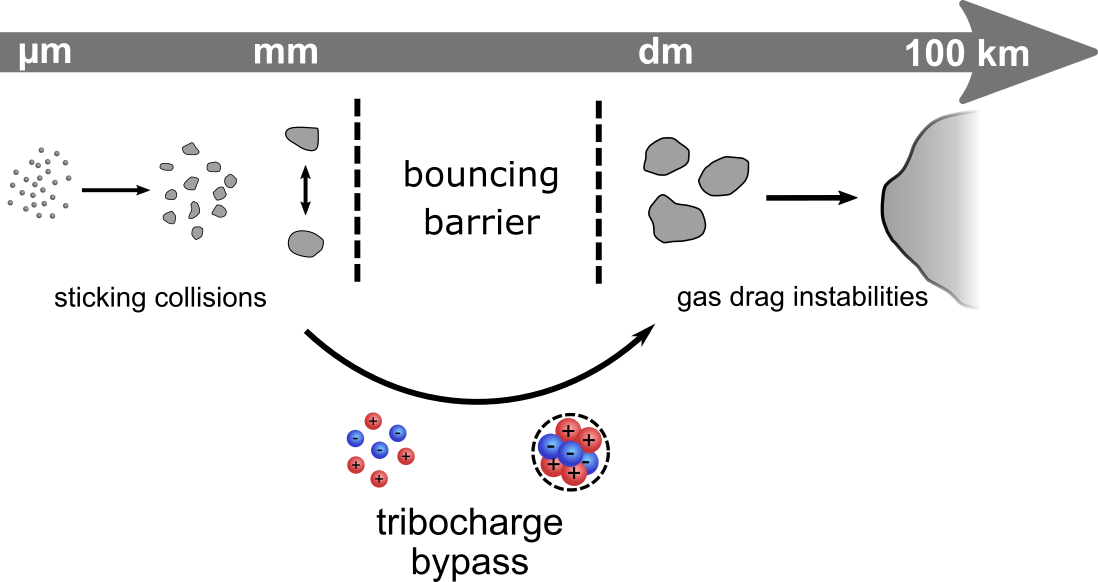}
	\caption{\label{bypass} The stability of tribocharged clusters can overcome the bouncing barrier in planetesimal formation.} 
\end{figure*}

\section{Methods}

\subsection{Experiments}

The experiment described here flew as an ESA payload on MASER 15 in November 2022 from Kiruna in northern Sweden. It was engineered by SSC (Swedish Space Corporation).
The experiment sketch is shown in the supplementary figure 1. Prior to launch, the experiment chamber was evacuated to $10^{-4}$ mbar. 
{The temperature of the experiment was between 274 K and 301 K during the whole procedure.}

The entire chamber can be shaken in a single direction. This was carried out for about 30 minutes prior to launch. The experiment commenced within minutes after the cessation of shaking. {Particles remain charged for the duration of the experiment then \citep{Becker2024_discharge}. The microgravity time can then be used in total, studying particle evolution.} The sample particles were still confined in a dedicated compartment during shaking, its walls coated with the same particles. The shaking induces collision between the grains, leading to tribocharging.  Despite their identical nature, the particles charge upon collision but without predictable preference, resulting in random net charges with positive or negative polarity and multipole charge patterns on their surface \citep{Paehtz2010, Yoshimatsu2017, Steinpilz2020, Steinpilz2020a, Onyeagusi2022}. 

The wall coating mitigates net charge bias due to different materials (basalt, metal, glass) or sizes (curvature wall - curvature spherical particle), ensuring an overall neutral particle cloud, as observed in prior experiments \citep{Steinpilz2020}. 
These and similar experiments with monodisperse glass and basalt particles of comparable size in more detail revealed exponentially distributed net charges on the grains, typically ranging between  $10^{6} .. 10^{8}$ elementary charges \citep{Haeberle2019, Steinpilz2020, Jungmann2022, Wurm2019}. These values serve as benchmarks for weakly and strongly charged clusters in the simulations discussed below. 

The experiment chamber underwent shaking with a large amplitude of 2.5\,mm and frequencies of 20\,Hz for several seconds. This accelerated some leftover particles near the walls initially, along the direction of shaking. These grains transferred momentum to others out of reach of the walls, heating them akin to a granular gas, and ultimately randomizing particle motion. 

\subsection{Numerical simulations}

We utilize the open source DEM package LIGGGHTS 3.8. We employed a dissipative Hertzian contact model (\texttt{gran model hertz} in the LIGGGHTS implementation), as described in \cite{kloss2012models} with linearized Johnson-Kendall-Roberts cohesion (\texttt{skrj2}). We generated a cluster of particles with the following properties. Particle diameter was set to $0.5\,$mm. The surface energy density in this model is set to $7 \cdot 10^4\,$ergs$/$cm$^3$, equivalent to a surface energy of $\sim 0.9\,$mJ$/$m$^2$ in the full JKR-contact. This value is on the lower end of reference values, to outline the effect of charge on granular dynamics. Charges were assumed to be homogeneously distributed on each grain, drawn from normal distributions with two different standard deviations $\sigma_w = 10^6\,e$ for the weakly charged and $\sigma_s=10^8\,e$ for the strongly charged example. 

To generate a compact cluster similar to the experiment but within a manageable simulation time, particles were dropped into a conical mesh with a flattened ground at $0.1\,$g of gravity. The mesh is vibrated to allow the particles to reposition according to their charge. After settling, gravity was set to zero and the mesh was removed. Further simulation parameters were a Youngs-Modulus of $5\cdot10^7\,$Pa, a Poisson ratio of $0.2$, a coefficient of restitution of $0.95$ and a coefficient of friction of $0.15$.
Regarding cohesion, increasing the cohesion energy density to $10^5\,$erg/cm$^3$, corresponding to a surface energy of $\sim 10\,$mJ/m$^2$, raises the limit for ejecta production above $0.5\,$m/s impactor velocities as well. 

Visual inspection revealed structural differences between weakly and strongly charged clusters, with the latter resembling experimental conditions more closely.

\section{Data availability}

The analyzed erosion sequence is provided as video \citep{Teiser2024}.

\section{Code availability}

The simulations are based on LIGGGHTS 3.8 (https://github.com/CFDEMproject/LIGGGHTS-PUBLIC.git).

\section{Acknowledgements}
We thank ESA for funding the suborbital flight as Project CHIP on MASER 15 and especially Astrid Orr and Philippe de Gieter for overseeing the project. 
This project received funding from DLR Space Administration with funds provided by the Federal Ministry for Economic Affairs and Climate Action (BMWK) under grant numbers  50 WM 2142 (STARK), 50 WM 2346 (CHAP) and 50 WK 2270C (AIMS-AIDEX).
We also thank the Swedish Space Corporation (SSC) for transferring the scientific requirements into this fabulous experiment. Special thanks go to K. Henriksson, C. Vesco, M. Lindh, M. Inga, K. Löth, S. Krämer, and A. Gierse.

The text of the manuscript was originally written by the authors in full. However, the language was edited by using ChatGPT. 

\section{Author contribution}

J.T. led the project. J.P. analyzed the experimental data. K.J. carried out the numerical simulations. F.C.O. carried out additional precursor experiments. J.E.K. advised in various phases of the project. D.D. worked on the design, construction and testing of the experiment hardware. G.W. participated in all phases of the project.

\section{Competing interests}

The authors declare no competing interests.


\begin{thebibliography}{10}
\expandafter\ifx\csname url\endcsname\relax
  \def\url#1{\burl{#1}}\fi
\expandafter\ifx\csname urlprefix\endcsname\relax\def\urlprefix{URL }\fi
\providecommand{\bibinfo}[2]{#2}
\providecommand{\eprint}[2][]{\url{#2}}
\providecommand{\doi}[1]{\url{https://doi.org/#1}}


\bibitem{Johansen2014}
\bibinfo{author}{{Johansen}, A.} \emph{et~al.}
\newblock \bibinfo{editor}{{Beuther}, H.}, \bibinfo{editor}{{Klessen}, R.~S.}, \bibinfo{editor}{{Dullemond}, C.~P.} \& \bibinfo{editor}{{Henning}, T.} (eds) \emph{\bibinfo{title}{{The Multifaceted Planetesimal Formation Process}}}.
\newblock (eds \bibinfo{editor}{{Beuther}, H.}, \bibinfo{editor}{{Klessen}, R.~S.}, \bibinfo{editor}{{Dullemond}, C.~P.} \& \bibinfo{editor}{{Henning}, T.}) \emph{\bibinfo{booktitle}{Protostars and Planets VI}}, \bibinfo{pages}{547} (\bibinfo{year}{2014}).
\newblock \eprint{1402.1344}.

\bibitem{Savvidou2023}
\bibinfo{author}{{Savvidou}, S.} \& \bibinfo{author}{{Bitsch}, B.}
\newblock \bibinfo{title}{{How to make giant planets via pebble accretion}}.
\newblock \emph{\bibinfo{journal}{Astron. Astrophys.}} \textbf{\bibinfo{volume}{679}}, \bibinfo{pages}{A42} (\bibinfo{year}{2023}).

\bibitem{Demirci2020accretion}
\bibinfo{author}{{Demirci}, T.} \& \bibinfo{author}{{Wurm}, G.}
\newblock \bibinfo{title}{{Accretion of eroding pebbles and planetesimals in planetary envelopes}}.
\newblock \emph{\bibinfo{journal}{Astron. Astrophys.}} \textbf{\bibinfo{volume}{641}}, \bibinfo{pages}{A99} (\bibinfo{year}{2020}).

\bibitem{Vazan2023}
\bibinfo{author}{{Vazan}, A.} \& \bibinfo{author}{{Ormel}, C.~W.}
\newblock \bibinfo{title}{{Rocky sub-Neptunes formed by pebble accretion: Rain of rock from polluted envelopes}}.
\newblock \emph{\bibinfo{journal}{Astron. Astrophys.}} \textbf{\bibinfo{volume}{676}}, \bibinfo{pages}{L8} (\bibinfo{year}{2023}).

\bibitem{Youdin2005}
\bibinfo{author}{Youdin, A.~N.} \& \bibinfo{author}{Goodman, J.}
\newblock \bibinfo{title}{Streaming instabilities in protoplanetary disks}.
\newblock \emph{\bibinfo{journal}{Astrophys. J.}} \textbf{\bibinfo{volume}{620}}, \bibinfo{pages}{459} (\bibinfo{year}{2005}).

\bibitem{Johansen2006}
\bibinfo{author}{{Johansen}, A.}, \bibinfo{author}{{Klahr}, H.} \& \bibinfo{author}{{Henning}, T.}
\newblock \bibinfo{title}{{Gravoturbulent Formation of Planetesimals}}.
\newblock \emph{\bibinfo{journal}{Astrophys. J.}} \textbf{\bibinfo{volume}{636}}, \bibinfo{pages}{1121--1134} (\bibinfo{year}{2006}).

\bibitem{Gerbig2020}
\bibinfo{author}{{Gerbig}, K.}, \bibinfo{author}{{Murray-Clay}, R.~A.}, \bibinfo{author}{{Klahr}, H.} \& \bibinfo{author}{{Baehr}, H.}
\newblock \bibinfo{title}{{Requirements for Gravitational Collapse in Planetesimal Formation{\textemdash}The Impact of Scales Set by Kelvin-Helmholtz and Nonlinear Streaming Instability}}.
\newblock \emph{\bibinfo{journal}{Astrophys. J.}} \textbf{\bibinfo{volume}{895}}, \bibinfo{pages}{91} (\bibinfo{year}{2020}).

\bibitem{Guettler2010}
\bibinfo{author}{G\"uttler, C.}, \bibinfo{author}{Blum, J.}, \bibinfo{author}{Zsom, A.}, \bibinfo{author}{Ormel, C.~W.} \& \bibinfo{author}{Dullemond, C.~P.}
\newblock \bibinfo{title}{The outcome of protoplanetary dust growth: pebbles, boulders, or planetesimals? - I. mapping the zoo of laboratory collision experiments}.
\newblock \emph{\bibinfo{journal}{Astron. Astrophys.}} \textbf{\bibinfo{volume}{513}}, \bibinfo{pages}{A56} (\bibinfo{year}{2010}).

\bibitem{Zsom2010}
\bibinfo{author}{{Zsom}, A.}, \bibinfo{author}{{Ormel}, C.~W.}, \bibinfo{author}{{G{\"u}ttler}, C.}, \bibinfo{author}{{Blum}, J.} \& \bibinfo{author}{{Dullemond}, C.~P.}
\newblock \bibinfo{title}{{The outcome of protoplanetary dust growth: pebbles, boulders, or planetesimals? II. Introducing the bouncing barrier}}.
\newblock \emph{\bibinfo{journal}{Astron. Astrophys.}} \textbf{\bibinfo{volume}{513}}, \bibinfo{pages}{A57} (\bibinfo{year}{2010}).

\bibitem{Wurm2021nat}
\bibinfo{author}{Wurm, G.} \& \bibinfo{author}{Teiser, J.}
\newblock \bibinfo{title}{Understanding planet formation using microgravity experiments}.
\newblock \emph{\bibinfo{journal}{Nat. Rev. Phys.}}  (\bibinfo{year}{2021}).


\bibitem{Birnstiel2011}
\bibinfo{author}{{Birnstiel}, T.}, \bibinfo{author}{{Ormel}, C.~W.} \& \bibinfo{author}{{Dullemond}, C.~P.}
\newblock \bibinfo{title}{{Dust size distributions in coagulation/fragmentation equilibrium: numerical solutions and analytical fits}}.
\newblock \emph{\bibinfo{journal}{Astron. Astrophys.}} \textbf{\bibinfo{volume}{525}}, \bibinfo{pages}{A11} (\bibinfo{year}{2011}).

\bibitem{Drazkowska2021}
\bibinfo{author}{{Drazkowska}, J.}, \bibinfo{author}{{Stammler}, S.~M.} \& \bibinfo{author}{{Birnstiel}, T.}
\newblock \bibinfo{title}{{How dust fragmentation may be beneficial to planetary growth by pebble accretion}}.
\newblock \emph{\bibinfo{journal}{Astron. Astrophys.}} \textbf{\bibinfo{volume}{647}}, \bibinfo{pages}{A15} (\bibinfo{year}{2021}).

\bibitem{Carrera2022}
\bibinfo{author}{{Carrera}, D.} \& \bibinfo{author}{{Simon}, J.~B.}
\newblock \bibinfo{title}{{The Streaming Instability Cannot Form Planetesimals from Millimeter-size Grains in Pressure Bumps}}.
\newblock \emph{\bibinfo{journal}{Astrophys. J.}} \textbf{\bibinfo{volume}{933}}, \bibinfo{pages}{L10} (\bibinfo{year}{2022}).

\bibitem{Drazkowska2014}
\bibinfo{author}{{Drazkowska}, J.} \& \bibinfo{author}{{Dullemond}, C.~P.}
\newblock \bibinfo{title}{{Can dust coagulation trigger streaming instability?}}
\newblock \emph{\bibinfo{journal}{Astron. Astrophys.}} \textbf{\bibinfo{volume}{572}}, \bibinfo{pages}{A78} (\bibinfo{year}{2014}).

\bibitem{Kelling2014}
\bibinfo{author}{Kelling, T.}, \bibinfo{author}{Wurm, G.} \& \bibinfo{author}{K{\"o}ster, M.}
\newblock \bibinfo{title}{Experimental study on bouncing barriers in protoplanetary disks}.
\newblock \emph{\bibinfo{journal}{Astrophys. J.}} \textbf{\bibinfo{volume}{783}}, \bibinfo{pages}{111} (\bibinfo{year}{2014}).


\bibitem{Kruss2016}
\bibinfo{author}{Kruss, M.}, \bibinfo{author}{Demirci, T.}, \bibinfo{author}{Koester, M.}, \bibinfo{author}{Kelling, T.} \& \bibinfo{author}{Wurm, G.}
\newblock \bibinfo{title}{Failed growth at the bouncing barrier in planetesimal formation}.
\newblock \emph{\bibinfo{journal}{Astrophys. J.}} \textbf{\bibinfo{volume}{827}}, \bibinfo{pages}{110} (\bibinfo{year}{2016}).

\bibitem{Demirci2019b}
\bibinfo{author}{{Demirci}, T.}, \bibinfo{author}{{Krause}, C.}, \bibinfo{author}{{Teiser}, J.} \& \bibinfo{author}{{Wurm}, G.}
\newblock \bibinfo{title}{{Onset of planet formation in the warm inner disk. Colliding dust aggregates at high temperatures}}.
\newblock \emph{\bibinfo{journal}{Astron. Astrophys.}} \textbf{\bibinfo{volume}{629}}, \bibinfo{pages}{A66} (\bibinfo{year}{2019}).

\bibitem{Arakawa2023}
\bibinfo{author}{{Arakawa}, S.} \emph{et~al.}
\newblock \bibinfo{title}{{Size Dependence of the Bouncing Barrier in Protoplanetary Dust Growth}}.
\newblock \emph{\bibinfo{journal}{Astrophys. J.l}} \textbf{\bibinfo{volume}{951}}, \bibinfo{pages}{L16} (\bibinfo{year}{2023}).

\bibitem{Pillich2021}
\bibinfo{author}{{Pillich}, C.}, \bibinfo{author}{{Bogdan}, T.}, \bibinfo{author}{{Landers}, J.}, \bibinfo{author}{{Wurm}, G.} \& \bibinfo{author}{{Wende}, H.}
\newblock \bibinfo{title}{{Drifting inwards in protoplanetary discs. II. The effect of water on sticking properties at increasing temperatures}}.
\newblock \emph{\bibinfo{journal}{Astron. Astrophys.}} \textbf{\bibinfo{volume}{652}}, \bibinfo{pages}{A106} (\bibinfo{year}{2021}).

\bibitem{Kruss2018}
\bibinfo{author}{{Kruss}, M.} \& \bibinfo{author}{{Wurm}, G.}
\newblock \bibinfo{title}{{Seeding the Formation of Mercurys: An Iron-sensitive Bouncing Barrier in Disk Magnetic Fields}}.
\newblock \emph{\bibinfo{journal}{Astrophys. J.}} \textbf{\bibinfo{volume}{869}}, \bibinfo{pages}{45} (\bibinfo{year}{2018}).

\bibitem{Bogdan2023}
\bibinfo{author}{{Bogdan}, T.}, \bibinfo{author}{{Pillich}, C.}, \bibinfo{author}{{Landers}, J.}, \bibinfo{author}{{Wende}, H.} \& \bibinfo{author}{{Wurm}, G.}
\newblock \bibinfo{title}{{The Curie line in protoplanetary disks and the formation of Mercury-like planets}}.
\newblock \emph{\bibinfo{journal}{Astron. Astrophys.}} \textbf{\bibinfo{volume}{670}}, \bibinfo{pages}{A6} (\bibinfo{year}{2023}).

\bibitem{Steinpilz2020}
\bibinfo{author}{{Steinpilz}, T.} \emph{et~al.}
\newblock \bibinfo{title}{{Electrical charging overcomes the bouncing barrier in planet formation}}.
\newblock \emph{\bibinfo{journal}{Nat. Phys.}} \textbf{\bibinfo{volume}{16}}, \bibinfo{pages}{225--229} (\bibinfo{year}{2020}).

\bibitem{Jungmann2018}
\bibinfo{author}{Jungmann, F.}, \bibinfo{author}{Steinpilz, T.}, \bibinfo{author}{Teiser, J.} \& \bibinfo{author}{Wurm, G.}
\newblock \bibinfo{title}{Sticking and restitution in collisions of charged sub-mm dielectric grains}.
\newblock \emph{\bibinfo{journal}{J. Phys. Comm.}}  (\bibinfo{year}{2018}).

\bibitem{Jungmann&Wurm2021}
\bibinfo{author}{Jungmann, F.} \& \bibinfo{author}{Wurm, G.}
\newblock \bibinfo{title}{Observation of bottom-up formation of charged grain aggregates related to pre-planetary evolution beyond the bouncing barrier}.
\newblock \emph{\bibinfo{journal}{Astron. Astrophys.}}  (\bibinfo{year}{2021}).

\bibitem{Teiser2021}
\bibinfo{author}{{Teiser}, J.}, \bibinfo{author}{{Kruss}, M.}, \bibinfo{author}{{Jungmann}, F.} \& \bibinfo{author}{{Wurm}, G.}
\newblock \bibinfo{title}{{A Smoking Gun for Planetesimal Formation: Charge-driven Growth into a New Size Range}}.
\newblock \emph{\bibinfo{journal}{Astrophys. J.}} \textbf{\bibinfo{volume}{908}}, \bibinfo{pages}{L22} (\bibinfo{year}{2021}).

\bibitem{Lee2015}
\bibinfo{author}{{Lee}, V.}, \bibinfo{author}{{Waitukaitis}, S.~R.}, \bibinfo{author}{{Miskin}, M.~Z.} \& \bibinfo{author}{{Jaeger}, H.~M.}
\newblock \bibinfo{title}{{Direct observation of particle interactions and clustering in charged granular streams}}.
\newblock \emph{\bibinfo{journal}{Nat. Phys.}} \textbf{\bibinfo{volume}{11}}, \bibinfo{pages}{733--737} (\bibinfo{year}{2015}).

\bibitem{Schneider2021}
\bibinfo{author}{{Schneider}, N.} \emph{et~al.}
\newblock \bibinfo{title}{{Experimental study of clusters in dense granular gas and implications for the particle stopping time in protoplanetary disks}}.
\newblock \emph{\bibinfo{journal}{\icarus}} \textbf{\bibinfo{volume}{360}}, \bibinfo{pages}{114307} (\bibinfo{year}{2021}).

\bibitem{Steinpilz2020a}
\bibinfo{author}{{Steinpilz}, T.}, \bibinfo{author}{{Jungmann}, F.}, \bibinfo{author}{{Joeris}, K.}, \bibinfo{author}{{Teiser}, J.} \& \bibinfo{author}{{Wurm}, G.}
\newblock \bibinfo{title}{{Measurements of dipole moments and a Q-patch model of collisionally charged grains}}.
\newblock \emph{\bibinfo{journal}{New J. Phys.}} \textbf{\bibinfo{volume}{22}}, \bibinfo{pages}{093025} (\bibinfo{year}{2020}).

\bibitem{Onyeagusi2022}
\bibinfo{author}{Onyeagusi, F.~C.}, \bibinfo{author}{Teiser, J.}, \bibinfo{author}{Schneider, N.} \& \bibinfo{author}{Wurm, G.}
\newblock \bibinfo{title}{Measuring electric dipole moments of trapped sub-mm particles}.
\newblock \emph{\bibinfo{journal}{J. Electrost.}} \textbf{\bibinfo{volume}{115}}, \bibinfo{pages}{103637} (\bibinfo{year}{2022}).


\bibitem{Weidenschilling1993}
\bibinfo{author}{{Weidenschilling}, S.~J.} \& \bibinfo{author}{{Cuzzi}, J.~N.}
\newblock \bibinfo{editor}{{Levy}, E.~H.} \& \bibinfo{editor}{{Lunine}, J.~I.} (eds) \emph{\bibinfo{title}{{Formation of Planetesimals in the Solar Nebula}}}.
\newblock (eds \bibinfo{editor}{{Levy}, E.~H.} \& \bibinfo{editor}{{Lunine}, J.~I.}) \emph{\bibinfo{booktitle}{Protostars and Planets III}}, \bibinfo{pages}{1031} (\bibinfo{year}{1993}).

\bibitem{Gurrutxaga2024}
\bibinfo{author}{{Gurrutxaga}, N.}, \bibinfo{author}{{Drazkowska}, J.} \& \bibinfo{author}{{Kleine}, T.}
\newblock \bibinfo{editor}{{Europlanet Society}} (ed.) \emph{\bibinfo{title}{{Modeling the formation of carbonaceous chondrite parent bodies}}}.
\newblock (ed.\bibinfo{editor}{{Europlanet Society}}) \emph{\bibinfo{booktitle}{European Planetary Science Congress}}, \bibinfo{pages}{EPSC2024--900} (\bibinfo{year}{2024}).

\bibitem{Krijt2015}
\bibinfo{author}{{Krijt}, S.}, \bibinfo{author}{{Ormel}, C.~W.}, \bibinfo{author}{{Dominik}, C.} \& \bibinfo{author}{{Tielens}, A.~G.~G.~M.}
\newblock \bibinfo{title}{{Erosion and the limits to planetesimal growth}}.
\newblock \emph{\bibinfo{journal}{Astron. Astrophys.}} \textbf{\bibinfo{volume}{574}}, \bibinfo{pages}{A83} (\bibinfo{year}{2015}).

\bibitem{Seizinger2013}
\bibinfo{author}{{Seizinger}, A.}, \bibinfo{author}{{Krijt}, S.} \& \bibinfo{author}{{Kley}, W.}
\newblock \bibinfo{title}{{Erosion of dust aggregates}}.
\newblock \emph{\bibinfo{journal}{Astron. Astrophys.}} \textbf{\bibinfo{volume}{560}}, \bibinfo{pages}{A45} (\bibinfo{year}{2013}).

\bibitem{Schraepler2018}
\bibinfo{author}{{Schr{\"a}pler}, R.}, \bibinfo{author}{{Blum}, J.}, \bibinfo{author}{{Krijt}, S.} \& \bibinfo{author}{{Raabe}, J.-H.}
\newblock \bibinfo{title}{{The Physics of Protoplanetary Dust Agglomerates. X. High-velocity Collisions between Small and Large Dust Agglomerates as a Growth Barrier}}.
\newblock \emph{\bibinfo{journal}{Astrophys. J.}} \textbf{\bibinfo{volume}{853}}, \bibinfo{pages}{74} (\bibinfo{year}{2018}).

\bibitem{Dominik2024}
\bibinfo{author}{Dominik, C.} \& \bibinfo{author}{Dullemond, C.~P.}
\newblock \bibinfo{title}{The bouncing barrier revisited: Impact on key planet formation processes and observational signatures}.
\newblock \emph{\bibinfo{journal}{Astron. Astrophys.}} \textbf{\bibinfo{volume}{682}}, \bibinfo{pages}{A144} (\bibinfo{year}{2024}).

\bibitem{Okuzumi2009}
\bibinfo{author}{Okuzumi, S.}, \bibinfo{author}{Tanaka, H.} \& \bibinfo{author}{Sakagami, M.-a.}
\newblock \bibinfo{title}{Numerical modeling of the coagulation and porosity evolution of dust aggregates}.
\newblock \emph{\bibinfo{journal}{Astrophys. J.}} \textbf{\bibinfo{volume}{707}}, \bibinfo{pages}{1247} (\bibinfo{year}{2009}).

\bibitem{Akimkin2020}
\bibinfo{author}{{Akimkin}, V.~V.}, \bibinfo{author}{{Ivlev}, A.~V.} \& \bibinfo{author}{{Caselli}, P.}
\newblock \bibinfo{title}{{Inhibited Coagulation of Micron-size Dust Due to the Electrostatic Barrier}}.
\newblock \emph{\bibinfo{journal}{Astrophys. J.}} \textbf{\bibinfo{volume}{889}}, \bibinfo{pages}{64} (\bibinfo{year}{2020}).

\bibitem{Becker2024_discharge}
\bibinfo{author}{{Becker}, T.} \emph{et~al.}
\newblock \bibinfo{title}{{Tribocharged solids in protoplanetary discs: internal and external discharge time-scales}}.
\newblock \emph{\bibinfo{journal}{Mon. Not. R. Astron. Soc.}} \textbf{\bibinfo{volume}{533}}, \bibinfo{pages}{413--422} (\bibinfo{year}{2024}).

\bibitem{Paehtz2010}
\bibinfo{author}{{P{\"a}htz}, T.}, \bibinfo{author}{{Herrmann}, H.~J.} \& \bibinfo{author}{{Shinbrot}, T.}
\newblock \bibinfo{title}{{Why do particle clouds generate electric charges?}}
\newblock \emph{\bibinfo{journal}{Nat. Phys.}} \textbf{\bibinfo{volume}{6}}, \bibinfo{pages}{364--368} (\bibinfo{year}{2010}).

\bibitem{Yoshimatsu2017}
\bibinfo{author}{{Yoshimatsu}, R.}, \bibinfo{author}{{Ara{\'u}jo}, N.~A.~M.}, \bibinfo{author}{{Wurm}, G.}, \bibinfo{author}{{Herrmann}, H.~J.} \& \bibinfo{author}{{Shinbrot}, T.}
\newblock \bibinfo{title}{{Self-charging of identical grains in the absence of an external field}}.
\newblock \emph{\bibinfo{journal}{Sci. Rep.}} \textbf{\bibinfo{volume}{7}}, \bibinfo{pages}{39996} (\bibinfo{year}{2017}).

\bibitem{Haeberle2019}
\bibinfo{author}{{Haeberle}, J.}, \bibinfo{author}{{Harju}, J.}, \bibinfo{author}{{Sperl}, M.} \& \bibinfo{author}{{Born}, P.}
\newblock \bibinfo{title}{{Granular ionic crystals in a small nutshell}}.
\newblock \emph{\bibinfo{journal}{Soft Mat.}} \textbf{\bibinfo{volume}{15}}, \bibinfo{pages}{7179--7186} (\bibinfo{year}{2019}).

\bibitem{Jungmann2022}
\bibinfo{author}{{Jungmann}, F.}, \bibinfo{author}{{Kruss}, M.}, \bibinfo{author}{{Teiser}, J.} \& \bibinfo{author}{{Wurm}, G.}
\newblock \bibinfo{title}{{Aggregation of sub-mm particles in strong electric fields under microgravity conditions}}.
\newblock \emph{\bibinfo{journal}{\icarus}} \textbf{\bibinfo{volume}{373}}, \bibinfo{pages}{114766} (\bibinfo{year}{2022}).

\bibitem{Wurm2019}
\bibinfo{author}{{Wurm}, G.}, \bibinfo{author}{{Schmidt}, L.}, \bibinfo{author}{{Steinpilz}, T.}, \bibinfo{author}{{Boden}, L.} \& \bibinfo{author}{{Teiser}, J.}
\newblock \bibinfo{title}{{A challenge for Martian lightning: Limits of collisional charging at low pressure}}.
\newblock \emph{\bibinfo{journal}{\icarus}} \textbf{\bibinfo{volume}{331}}, \bibinfo{pages}{103--109} (\bibinfo{year}{2019}).

\bibitem{kloss2012models}
\bibinfo{author}{Kloss, C.}, \bibinfo{author}{Goniva, C.}, \bibinfo{author}{Hager, A.}, \bibinfo{author}{Amberger, S.} \& \bibinfo{author}{Pirker, S.}
\newblock \bibinfo{title}{Models, algorithms and validation for opensource dem and cfd--dem}.
\newblock \emph{\bibinfo{journal}{Prog. Comp. Fluid Dyn.}} \textbf{\bibinfo{volume}{12}}, \bibinfo{pages}{140--152} (\bibinfo{year}{2012}).

\bibitem{Teiser2024}
\bibinfo{author}{{Teiser}, J.} \emph{et~al.}
\newblock \bibinfo{title}{{Growing Super-Large Pre-Planetary Pebbles to an Impact Erosion Limit}}.
\newblock \emph{\bibinfo{journal}{figshare Dataset}}  (\bibinfo{year}{2024}). 
\newblock
\doi{10.6084/m9.figshare.27210705}

\end{thebibliography}
\end{document}